\documentclass[aps,prd,superscriptaddress,amsfonts,amssymb,amsmath,nofootinbib,showpacs,showkeys]{revtex4}
\usepackage{ulem,color,wrapfig,graphicx}

\usepackage{amsfonts,amssymb,amsmath,amsthm,mathrsfs}
\usepackage{url}
\newtheorem{thrm}{Theorem}[section]
\newtheorem{lem}[thrm]{Lemma}

\newtheorem{cor}[thrm]{Corollary}
\newtheorem{remark}[thrm]{Remark}
\theoremstyle{definition}
\newtheorem{definition}[thrm]{Definition}

\newtheorem{example}[thrm]{Example}

\begin{document}

\author{Salvatore Capozziello}
\email{capozziello@na.infn.it}
\address{Dipartimento di Fisica ``E. Pancini", Universit\`a di Napoli {\it Federico II}, Napoli,\\ and
INFN Sez. di Napoli, Compl. Univ. di Monte S. Angelo, Edificio G, Via Cinthia, I-80126, Napoli, Italy,}
\address{Laboratory for Theoretical Cosmology, Tomsk State University of Control Systems and Radioelectronics (TUSUR), 
\\634050 Tomsk (Russia).}

\author{Carlo Alberto Mantica}
\email{carlo.mantica@mi.infn.it}
\address{I.I.S. Lagrange, Via L. Modignani 65, 
I-20161, Milano, Italy \\
and INFN sez. di Milano,
Via Celoria 16, I-20133 Milano, Italy}

\author{Luca Guido Molinari}
\email{luca.molinari@unimi.it}
\address{Dipartimento di Fisica ``A. Pontremoli'',
Universit\`a degli Studi di Milano\\ and INFN sez. di Milano,
Via Celoria 16, I-20133 Milano, Italy.}
\email{luca.molinari@unimi.it}


\title[Cosmological perfect fluids in higher-order gravity]{Cosmological perfect fluids  in higher-order gravity}

\begin{abstract}
We prove that in Robertson-Walker space-times (and in generalized Robertson-Walker spacetimes of dimension greater than 3 with divergence-free Weyl tensor) all higher-order gravitational corrections of the Hilbert-Einstein Lagrangian density $F(R,\square R, ... , \square^k R)$ have the form of perfect fluids in the 
field equations. This statement  definitively allows to deal  with dark energy fluids as  curvature effects.
\end{abstract}
\maketitle

\date{\today}
\maketitle

\section{Introduction} 
The guiding principles for the Einstein equations are covariance and the Newtonian limit, and the 
gravitational action has the simplest conceivable form: the invariant space-time integral of the curvature scalar $R$. 
Despite being tested in a wide range of cases, the question arises whether the expression is an approximation of a more fundamental one or it is the final theory of gravity working from ultraviolet (quantum gravity) up to infrared scales (astrophysics and cosmology).  There are various motivations for exploring extensions of gravity: its possible origin from string theory, quantization of gravity, a mechanism for the early inflation, the quest for a geometric origin of the ``dark side'' of the universe, or simply a mathematical question of principle (what happens if we modify the Hilbert-Einstein action because there is no {\it a priori} reason to investigate the simplest one). They are surveyed in several review papers \cite{CapozDeL,CF08,Nojiri17,Zerbini, BCNO12}. 

Among the extensions, the $F(R)$ and $F(R,G)$ theories are the most studied. The first one replaces $R$ in the action with a function of it \cite{quintessence}, in the other the Lagrangian is a function of $R$ and of the Gauss-Bonnet scalar 
$G= R_{ijkl}R^{ijkl} -4 R_{ij}R^{ij} + R^2$ \cite{Odintsov}. Lovelock's action contains scalars obtained from products 
of the Riemann tensor  \cite{Lovelock}. In general, higher-order gravity  can be characterized by an action with terms $\square^m R$ (see \cite{Schmidt} for details). \\
In applications to large-scale cosmology, the constraint of spacetime homogeneity and isotropy imposes the Robertson-Walker metric as the background characterizing, in principle,  any   modified or extended theory of gravity.\\
For $F(R)$ models,  it is possible to  prove that each new (Hessian) term in the Einstein equations has the perfect fluid form for any smooth function $F$ provided that the space-time is a generalized Robertson-Walker (GRW) space-time with divergence-free Weyl tensor \cite{CMM_fR}. In a GRW space-time of dimension $n=4$, the condition $\nabla^m C_{jklm}=0$ is equivalent to $C_{jklm}=0$ and the space-time is Robertson-Walker (RW) (see Prop.4.1 in \cite{MaMoJMP16}).\\
The result can be extended to $F(R,G)$ theories \cite{CMM_fRG} in RW space-times: the new terms have the perfect fluid
form, i.e. they contribute to the energy-momentum tensor of matter $T_{ij}=(p+\mu) u_iu_j+pg_{ij}$ as terms 
that modify the total pressure and energy density or, in other words, can be dealt as a new effective fluid.\\ 
In this work we consider higher-order gravity with action \cite{Schmidt,Wands,CapozDeL}
\begin{align}
{\cal A}=\frac{1}{2\kappa} \int d^n x \sqrt{-g} \, F(R,\square R, ...,\square^k R) + {\cal A}^{(m)}
\end{align}
where $F$ is a smooth function of $k+1$ real variables, and ${\cal A}^{(m)}$ is
the standard action for matter fields. The condition of minimum action yields the  field equations \cite{Schmidt}, where the higher-order geometric terms can be collected in the right-hand side: 
\begin{align}
&R_{ij}-\tfrac{1}{2}Rg_{ij} =\kappa T^{(m)}_{ij} + \kappa T_{ij}^{(HG)} \label{Einsteq}\\
&\kappa T_{ij}^{(HG)}= (1- \Theta_0) R_{ij} +\tfrac{1}{2}(F-R-2\square \Theta_0) g_{ij} 
+\nabla_i\nabla_j \Theta_0  \label{fieldeq}\\
&-\tfrac{1}{2}\sum_{a=1}^k [g_{ij} \nabla_l
(\Theta_a\nabla^l (\square^{a-1} R)) -(\nabla_i\Theta_a) (\nabla_j \square^{a-1} R)  - (\nabla_j \Theta_a )( \nabla_i \square^{a-1} R)] \nonumber
\nonumber \\
&\Theta_a = \sum_{b=a}^k \square^{b-a} \frac{\partial F}{\partial (\square^b R)}, \quad a=0,...,k. 
\end{align}
Remarkably, the higher-order gravity contribution has the property $\nabla^i T^{(HG)}_{ij} =0$ \cite{CapozDeL,Schmidt}.
This means that the contracted Bianchi identities guarantee conservation laws and then the main property of any perfect fluid.

In analogy to the findings for $F(R)$ and $F(R,G)$ gravity, it is possible to prove
that also for higher-order gravity, the geometric corrections to the Einstein field equations have the perfect fluid form according to the following theorem:
\begin{thrm}
In a generalized Robertson-Walker space-time with $\nabla^m C_{jklm}=0$, 
the higher-order gravity tensor $T^{(HG)}_{ij}$ has the form of a perfect fluid with velocity field $u_i$:
\begin{align}
\kappa T^{(HG)}_{ij} =&
u_iu_j \left [ (1- \Theta_0) \frac{R-n\xi}{n-1}  + \ddot\Theta_0 -\varphi \dot \Theta_0 + \sum_{a=1..k}
\dot\Theta_a (u^l\nabla_l \square^{a-1} R) \right] \label{pfexpr}\\
& + g_{ij}  \left [ (1-\Theta_0)  \frac{R-\xi}{n-1}  +\tfrac{1}{2}(F-R) - \square \Theta_0  -\varphi \dot\Theta_0  
-\tfrac{1}{2}\sum_{a=1..k}  \nabla_l (\Theta_a\nabla^l (\square^{a-1} R) )    \right ]. \nonumber 
\end{align}
\end{thrm}
The Ricci tensor is given in eq.\eqref{Ricci_GRW}. By writing 
\begin{equation}
T^{HG}_{ij}= \kappa (p^{HG}+\mu^{HG} )u_iu_j + \kappa p^{HG}g_{ij}\,,
\end{equation}
and with the simplifications offered by Lemma \ref{lem1}, \ref{lem2} and Corollary \ref{cor1}, 
we obtain the energy density and pressure contributions of the higher-order geometry,  in the co-moving frame $u^0=1, u^i=0$: 
\begin{align}
\kappa\mu^{HG}=&-\tfrac{1}{2}(F-R)+\xi (\Theta_0-1)-(n-1)\varphi \dot\Theta_0\\
& +\tfrac{1}{2}{\sum}_{a=1}^k [\dot\Theta_a u^l\nabla_l (\square^{a-1} R) + \Theta_a \square^a R], \nonumber\\
\kappa p^{HG}= &\tfrac{1}{2}(F-R) + (1-\Theta_0)  \frac{R-\xi}{n-1}  +\ddot \Theta_0  +(n-2)\varphi \dot\Theta_0  \\
&+\tfrac{1}{2}{\sum}_{a=1}^k  [\dot \Theta_a u^l\nabla_l  (\square^{a-1} R) - \Theta_a \square^a R]. \nonumber
\end{align}

The result is achieved by exploiting the special properties of scalars whose variation in GRW space-time is geodesic,
$\nabla_k S=-u_k\dot S$. The scalar curvature $R$ has such variation if we also require 
$\nabla^m C_{jklm}=0$. Such properties are obtained by a covariant characterization of GRW space-times, 
that includes RW as a special case. The first characterization was obtained by Bang-Yen Chen in 2014 \cite{Chen14}.
It was found equivalent to the existence of a velocity vector field that is shear-free, vorticity-free and geodesic \cite{[29]}, that
describes the cosmological fluid. 

Section 2 is devoted to a brief presentation of GRW space-times, and to
the properties of the so-called {\it perfect scalars}, forming a closed algebra for the action of the operator $\square =g^{jk}\nabla_j\nabla_k$. In Section 3 we prove the theorem and, in Section 4, we discuss some examples. Discussion and conclusions are given in Section 5.

\section{GRW space-times and perfect scalars}
 A generalized Robertson-Walker (GRW) space-time (see \cite{GRWSurv17} for a survey) is characterized by the warped metric 
\begin{align}
ds^2 = -(dt)^2+a(t)^2 g^*_{\mu\nu} dx^\mu dx^\nu \label{warped}
\end{align}
where $g^*$ is the positive-definite metric of a Riemannian sub-manifold, $a(t)$ is the scale parameter. Here $H=\dot a/a$ is the Hubble parameter. A RW space-time is the special case where the sub-manifold is isotropic and homogeneous or, equivalently, the Weyl tensor $C_{jklm}$ vanishes.\\ 
It is covariantly characterized by the existence of a velocity field $u_i u^i=-1$ that is
shear-free, vorticity-free and geodesic \cite{[29]}:
\begin{align}
\nabla_i u_j = \varphi \, h_{ij}
\end{align}
where $h_{ij}=g_{ij}+u_iu_j$ is a projection and $\nabla_i \varphi = -u_i \dot \varphi $. A dot means a derivative $u^k\nabla_k$.
In the comoving frame, the metric takes the warped form \eqref{warped}, $\varphi $ is the Hubble parameter, and a dot is a time derivative.\\
If we also require that the Weyl tensor of the GRW space-time is harmonic, $\nabla_m C_{jkl}{}^m=0$, then the Ricci tensor has the perfect fluid form \cite{MaMoJMP16}:
\begin{align}
 R_{jk} = &\frac{R-n\xi}{n-1} u_ju_k + \frac{R-\xi}{n-1}  g_{jk} \label{Ricci_GRW}
 \end{align}
where $R$ is the curvature scalar and $\xi =(n-1)(\dot\varphi +\varphi^2)$ is the eigenvalue $R_{ij}u^j=\xi u_i$.

We now introduce a useful novel concept:

\begin{definition}
A scalar $S$ is ``perfect'' if $\nabla_i S= -u_i \dot S$, where $\dot S=u^k\nabla_k S$.
\end{definition}
\begin{lem}\label{lem1} 
If a scalar is perfect, then $\dot S$ is perfect, i.e. $\nabla_j \dot S =-u_j \ddot S$.
\begin{proof}
$\nabla_i \dot S = \nabla_i u_k\nabla^k S = \varphi h_{ik} \nabla^k S + u^k\nabla_i\nabla_k S$. The first term is zero
because $S$ is perfect. In the second term, let us  exchange derivatives and obtain $u^k\nabla_k (\nabla_i S) =
-u^k\nabla_k (u_i\dot S) = -u_i u^k\nabla_k \dot S$. Then: $\nabla_i \dot S = -u_i \ddot S$ i.e. $\dot S$ is perfect too.
\end{proof}
\end{lem}
If $S$ is perfect then $\dot S, \ddot S, ...$ are perfect. The sum and the product of perfect scalars are perfect.
Since $\varphi $ is perfect in a GRW (by definition), then $\varphi^2$ and $\dot\varphi $ are perfect, and their combination 
$\xi $ is perfect.

\begin{lem}\label{lem2}
If a scalar field $S$ is perfect then the Hessian $\nabla_i\nabla_j S$ has the perfect-fluid form:
\begin{align}
\nabla_i\nabla_j S= -\varphi g_{ij} \dot S+u_iu_j(\ddot S -\varphi \dot S)  \label{Spf}
\end{align}
\begin{proof}
$\nabla_i\nabla_j S = \nabla_i(-u_j \dot S) = -\varphi h_{ij} \dot S -u_j \nabla_i \dot S = -\varphi h_{ij} \dot S +u_iu_j 
\ddot S $.
\end{proof}
\end{lem}

\begin{cor}\label{cor1}
If a scalar field $S$ is perfect, then $\square S = -(n-1)\varphi \dot S -\ddot S $ is perfect, being a combination of
perfect fields, and the scalar fields $\square^a S$, $a=1,2,...$  are perfect.
\end{cor}

\begin{lem}
In a GRW space-time with perfect fluid Ricci tensor, $R$ is a perfect scalar and 
\begin{align}
\dot R - 2\dot\xi = - 2\varphi (R-n\xi) \label{dotR} 
\end{align}
\begin{proof}
The Bianchi identity $\nabla^k R_{kj}=\frac{1}{2}\nabla_j R$ is evaluated with the Ricci tensor \eqref{Ricci_GRW}:
\begin{align*}
 \tfrac{1}{2}(n-1)\nabla_j R =& \nabla^k [(R-n\xi) h_{kj}] +(n-1)\nabla_j\xi\\
  =&\nabla_j R -\nabla_j \xi + u_j(\dot R -n\dot \xi )+ (R-n\xi) (n-1) \varphi u_j \\
  =& \nabla_j R +u_j \dot R +u_j [\varphi (R-n\xi) -\dot\xi ](n-1)
\end{align*}
Contraction with $u^j$ gives \eqref{dotR} and simplification of the last equation gives $\nabla_j R =-u_j \dot R$.
\end{proof}
\end{lem}
We now have all the ingredients to prove the main theorem.

\section{Proof of the theorem}
Consider a real smooth function of $(k+1)$ real variables $F(y_0,y_1,\ldots , y_k)$ and the corresponding 
$F(R,\square R, \square^2 R, \dots ,\square^k R)$, where each argument $y_a=\square^a R$ is a perfect scalar. \\
- $F$, as a function of $(R,\square R, ...,\square^k R)$, is a perfect scalar:
$$\nabla_j F = \sum_{a=0}^n \frac{\partial F}{\partial  y_a} \nabla_j (\square^a R) = -u_j \sum_{a=0}^n \frac{\partial F}{\partial  y_a} u^l\nabla_l (\square^a R) = -u_j (u^l\nabla_l F) $$
- $F^a \equiv (\partial F/\partial y_a)$, as a function of $(R,\square R, ...,\square^k R)$, is a perfect scalar:
$$  \nabla_j F^a = \sum_{b=1..k}  \frac{\partial F^a}{\partial y_b} \nabla_j (\square^b R)
= -u_j  u^l \sum_{b=1..k}  \frac{\partial F^a}{\partial y_b} \nabla_l (\square^b R) = -u_j (u^l\nabla_l F^a )$$
As a consequence, also $\square^c F^a$ is perfect for any power $c$, as well as the linear combinations 
$\Theta_a =  \sum_{b=a}^k \square^{b-a} F^b$, $a=0,1,...,k$.\\
By means of \eqref{Spf} and by the properties of perfect scalars, one obtains that
in a GRW space-time with harmonic Weyl tensor, the field equations \eqref{fieldeq} gain the
form of the perfect fluid, eq.\eqref{pfexpr}. {\it Q.E.D.}

\begin{remark}
In \cite{CMM_fR} we proved that a conformal transformation $\bar g =e^{2\sigma } g$ with $\nabla_k\sigma =-u_k\dot\sigma $ (i.e. $\sigma $ is a perfect scalar) maps a GRW space-time $\{{\cal M},g\}$
to a GRW space-time $\{{\cal M},\bar g\}$ and that the condition $\nabla^m C_{jklm}=0$ transforms into $\bar\nabla 
\bar C_{jklm}=0$. Here $\cal M$ is the manifold and $g$ is the metric. Then if the Ricci tensor in $\{{\cal M},g\}$ is a perfect-fluid tensor, so it is $\bar R_{ij}$.\\
In higher-order gravity, for harmonic GRW space-time, if $\Theta_0>0$, let $ \sigma =\frac{1}{n-2}\log \Theta_0$. The conformal map takes the higher-order gravity (defined in the Jordan frame) to a theory (defined in the Einstein frame) with fields $\phi_m =\square^m R $, $m=0,1,...,k$ that are minimally coupled to the ordinary Einstein gravity (see also \cite{Gott}).
\end{remark}

\section{higher-order gravity in RW space-time}
We give some example of the above results. 
Let us specialize to dimension $n=4$, where the space-time is Robertson-Walker. Being all terms in
the Einstein equation \eqref{Einsteq} proportional to $ g_{ij} $ or $u_i u_j $, the equation splits into two coupled
ones, that may be written as follows:
\begin{gather}
 R-2\xi  = 2 \kappa (\mu^m +\mu^{HG} ) \label{E1} \\
 R -4\xi  = 3\kappa ( p^m+p^{HG}+\mu^m+\mu^{HG} ) \label{E2}
\end{gather}
By inserting the expressions of $\mu^{HG}$ and $p^{HG}$ into the above cosmological equations, one obtains two non-linear higher-order differential equations for $R(t)$ and $\varphi (t)=\dot a/a$, to be considered with the equations of state for the matter terms  in $\mu^m $ and $p^m$.
\begin{remark}
Eq.\eqref{dotR} is equivalent to one of the two Einstein equations. \\
First, recall that each term $T_{ij}$ in the right-hand side of the Einstein equation \eqref{Einsteq} has zero divergence, 
and has the perfect fluid structure $T_{ij}=(p+\mu) u_iu_j + p g_{ij}$. The equation $\nabla^i T_{ij}=0$ is
$(\dot p+\dot \mu) u_j + \varphi (n-1) (p+\mu) u_j  + \nabla_j p=0$ i.e. 
$$ \nabla_j p=-u_j \dot p, \qquad \dot \mu = - 3\varphi (p+\mu). $$
The dot of eq.\eqref{E1}, use of the above property and \eqref{E2} give: $\dot R-2\dot\xi = 
2\kappa (\dot\mu^m +\dot \mu^{HG}) = -6\kappa \varphi (p^m+p^{HG}+\mu^m+\mu^{HG}) = -2\varphi (R-4\xi )$. \\%
The solution in $n=4$ of eq.\eqref{dotR} is
\begin{align}
 R(t)=2\xi + (R^*/a^2 ) + 6\varphi^2 = 12\varphi^2 +6\dot\varphi +(R^*/a^2)  \label{Rsol}
 \end{align}
where $R^*$ is the constant curvature of the space. 
\end{remark}

\begin{example}[Sixth-order gravity]{\quad}\\
As an illustration, we consider the case $F(R,\square R)=R+\gamma R\square R$. It is 
$\Theta_0=1+2\gamma \square R$, $\Theta_1=\gamma R$. The Einstein  equations \eqref{fieldeq}
involve sixth order derivatives of the metric tensor:
\begin{align}
R_{ij}-\tfrac{1}{2} R g_{ij} = \kappa T_{ij}^{(m)} + 2\gamma [ \nabla_i\nabla_j\square R  - R_{ij}\square R 
+\tfrac{1}{2}(\nabla_i R )(\nabla_j R )]  \label{sixth} \\
-2\gamma g_{ij} [\square^2 R +\tfrac{1}{4} (\nabla_k R)(\nabla^k R)]  \nonumber 
\end{align}
They coincide with eq. 2.2 in the paper by Gottl\"ober et al. \cite{Gott}. The most general sixth-order terms should  also
include
cubic products of the Ricci scalar, the Ricci and Riemann tensors, as well as terms 
like $R_{ij}\square R^{ij}$ \cite{Folacci}.\\
By restricting the solution of \eqref{sixth} to be a RW metric of dimension $n=4$, the theorem provides the perfect fluid parameters
of sixth order gravity:
\begin{align*}
\kappa\mu^{VI}=& 2\xi \gamma \square R - 6\gamma  \varphi \dot\square R +\tfrac{1}{2} \gamma (\dot R)^2 \nonumber\\
=&-18 \varphi^3\gamma \dot R  - 6\dot\varphi \gamma \ddot R
+12\gamma \varphi^2 \ddot R+6\gamma \varphi \dddot R  +\tfrac{1}{2} \gamma (\dot R)^2 \nonumber\\
\kappa p^{VI}= & -\tfrac{2}{3}\gamma  (R-\xi) \square R  +2\gamma \ddot \square R  + 4\gamma \varphi \dot\square R  +\tfrac{1}{2}  
\gamma (\dot R )^2 \nonumber\\
=&    -6\gamma(\varphi^3+3\varphi\dot\varphi+\ddot\varphi)\dot R +2\gamma\varphi R\dot R
+\tfrac{1}{2}\gamma (\dot R)^2\nonumber \\
&+\tfrac{2}{3}\gamma R\ddot R-14\gamma (\dot\varphi+\varphi^2) \ddot R -10\gamma\varphi\dddot R -2\gamma \ddddot R 
\end{align*}
The Einstein eq. \eqref{E1} {\it in vacuo} is then:
\begin{align}
R-2\xi = -36 \varphi^3\gamma \dot R  - 12\gamma (\dot\varphi -2\varphi^2)\ddot R
+12\gamma \varphi \dddot R  + \gamma (\dot R)^2\,.
\end{align} 
Considering eq.\eqref{Rsol},  it becomes a non-linear equation for the Hubble parameter $\varphi $.
\end{example}

\section{Discussion and Conclusions}

As discussed in \cite{Cardo}, dark energy and dark matter can be seen as curvature effects induced in the large scale, infrared limit of gravitational interaction. However, this statement is phenomenological and has to be compared with observations. In particular, cosmography can be an efficient tool to discriminate among concurring  models as discussed, for example, in  \cite{OrlandoDunsby,ester1,ester2,revcosmo}. 

Here we have rigorously demonstrated that, in the specific case of higher-order gravity, further terms entering the Hilbert-Einstein action can be dealt with the standard of perfect fluid, in the cosmological context of GRW metrics. The key ingredient to achieve this result is the fact that curvature terms involving $\square$ operators can be reduced to perfect scalars. Besides the mathematical result, the approach is interesting because it allows to point out the fundamental origin of scalar fields entering at early times (through inflation) and at late times (through dark energy) into the cosmological evolution. In both cases, as firstly pointed out by Starobinsky \cite{starobinsky, barrow}, they could have a geometric origin related to the formulation of  quantum field theory in curved spacetimes \cite{birrell}.

\section*{Acknowledgments}
S.~C. acknowledges the support of  INFN ({\it iniziative specifiche} TEONGRAV and QGSKY).
This paper is based upon work from COST action CA15117 (CANTATA), supported by COST (European 
Cooperation in Science and Technology).

\end{document}